\documentclass[12pt,a4paper]{article}

\usepackage[utf8]{inputenc}
\usepackage[T1]{fontenc}
\usepackage[english]{babel}
\usepackage{amsmath,amssymb}
\usepackage{graphicx}
\usepackage{hyperref}
\usepackage{natbib}
\usepackage{geometry}
\usepackage{tabularx}
\usepackage{float}
\usepackage{booktabs}
\usepackage{longtable}
\usepackage{array}
\usepackage{multirow}
\usepackage{tikz}
\usepackage{pgfplots}
\pgfplotsset{compat=1.18}
\usetikzlibrary{trees,shapes,arrows,positioning,shadows,calc,fit,backgrounds}

\hypersetup{
    colorlinks=true,
    linkcolor=blue,
    citecolor=blue,
    urlcolor=cyan
}

\definecolor{aiblue}{RGB}{52,152,219}
\definecolor{mlgreen}{RGB}{46,204,113}
\definecolor{dlpurple}{RGB}{155,89,182}
\definecolor{rlorange}{RGB}{230,126,34}
\definecolor{nlpred}{RGB}{231,76,60}
\definecolor{optgray}{RGB}{149,165,166}
\definecolor{bigdatateal}{RGB}{26,188,156}
\definecolor{esggreen}{RGB}{39,174,96}
\definecolor{finblue}{RGB}{41,128,185}

\title{\textbf{AI-Powered Sustainable Finance: An Integrative Taxonomy and Framework of AI Applications for Sustainable Investment Decision-Making}}

\author{
    Eduardo C. Garrido-Merch\'an\thanks{Corresponding author: ecgarrido@comillas.edu} \\[0.2em]
    \normalsize Instituto de Investigaci\'on Tecnol\'ogica (IIT), \\
    \normalsize Universidad Pontificia Comillas, Madrid, Spain
    \and
    Esther Vaquero Lafuente \\[0.2em]
    \normalsize Instituto de Investigaci\'on Tecnol\'ogica (IIT), \\
    \normalsize Universidad Pontificia Comillas, Madrid, Spain
    \and
    Elisa Aracil \\[0.2em]
    \normalsize Instituto de Investigaci\'on Tecnol\'ogica (IIT), \\
    \normalsize Universidad Pontificia Comillas, Madrid, Spain
}

\date{\today}

\begin{document}

\maketitle

\begin{abstract}
The integration of Artificial Intelligence into sustainable finance represents a transformative paradigm shift in how Environmental, Social, and Governance factors are analyzed, predicted, and incorporated into investment decisions. This review provides a comprehensive taxonomy of AI approaches applicable to sustainable investment decision-making, categorizing methodologies based on their underlying algorithms and their impact on ESG-related financial processes. The proposed AI Taxonomy includes machine learning paradigms---including supervised, unsupervised, and reinforcement learning---as well as natural language processing techniques and optimization algorithms, examining their specific applications in ESG score prediction, controversy detection, portfolio management, and sustainability report analysis. By synthesizing findings from the recent literature, a framework emerges on AI-powered sustainable finance that identifies technological applications to overcome ESG data barriers.
\end{abstract}


\section{Introduction}

Sustainable finance directs capital toward activities that foster sustainable development and long-term productivity while addressing environmental and social challenges \citep{porter1995, schoenmaker2017, oecd2020}. Consequently, Environmental, Social, and Governance (ESG) criteria have become central to investment decision-making, with investors seeking to align financial returns with positive societal impact \citep{friede2015esg, gillan2021, eccles2014}. This shift is reflected in the systematic integration of sustainability considerations into investment criteria \citep{oecd2020} and in the rapid expansion of the market for green, social, sustainability, and sustainability-linked (GSSS) financial instruments \citep{bis2023, gsia2021, worldbank2022}. As a result, the global sustainable investment market has grown exponentially in recent years, with ESG-integrated assets under management reaching trillions of dollars worldwide \citep{gsia2022}, signaling a fundamental transformation in how the financial industry perceives the relationship between sustainability and financial performance. The rise of sustainable finance reinforces the role of finance for advancing the common good and accelerating the sustainability transition, which demands massive capital mobilization \citep{schoenmaker2019, un2012, un2015} through carbon finance, climate finance, green finance, transition finance, biodiversity finance, and social finance among other \citep{unep2016}.

Nevertheless, significant barriers impede filling the global sustainable finance investment gap \citep{unctad2023}. These challenges stem primarily from difficulties in assessing sustainability performance, including conceptual ambiguity in ESG metrics, self-reported bias, insufficient disclosure standards, and regulatory fragmentation \citep{berg2022aggregate}. The resulting information asymmetries undermine trust—central to financial markets \citep{rajan2003, xu2020}, and are exacerbated by concerns over greenwashing and impression management, which complicate sustainability risk assessment and weaken confidence among market participants \citep{boiral2017, redondo2024}. At the same time, the growing use of alternative data sources such as social media posts and satellite imagery \citep{goodell2021} introduces additional complexity due to the dispersion, scale and unstructured nature of sustainability-related information.  

The rapid advancement of Artificial Intelligence (AI) technologies has opened unprecedented opportunities to mitigate these hurdles by enabling the integration, processing, and analysis of vast volumes of structured and unstructured data \citep{goodell2021}, identify complex patterns, and generate actionable insights that would be impossible for human analysts to extract manually \citep{alsartawi2023}. This capability is particularly valuable in the ESG domain, where AI enhances the objectivity, transparency, and predictive quality of ESG information \citep{bis2023}, thereby supporting more reliable sustainability risk assessment and investment decision-making \citep{alsartawi2022}. 

However, despite the growing interest in applying AI to sustainable finance \citep{elbouknify2026, eti2026, martinmelero2025, souto2026}, the field lacks a comprehensive and structured overview on how AI can foster sustainable finance particularly with respect to the  available AI technologies and their specific applications. Existing literature often focuses on individual techniques or narrow application domains \citep{goodell2021} making it challenging for researchers and practitioners to understand the full landscape of possibilities. This review aims to address this gap by providing a comprehensive taxonomy of AI approaches applicable to sustainable investment decision-making, a systematic review of existing applications categorized by AI technology and ESG domain, an analysis of current challenges and limitations, and identification of promising directions for future research.  A framework for AI powered sustainable finance emerges, that facilitates scholars an overview of extant research, identify gaps, and provide a structure for future investigation to improve the empirical and conceptual basis of sustainable finance. Through our proposed framework we provide a structured approach to realize AI’s full potential for sustainable finance, supporting practitioners’ integration into mainstream financial practices, regulatory progress, and knowledge advancements.

This paper contributes to the growing literature on AI in sustainable finance and ESG investing \citep{alsartawi2022}, which examines how data-driven methods can support sustainability-related financial decision-making by offering a comprehensive taxonomy of AI approaches across ESG applications. In addition, the paper contributes to the literature about quality and credibility of sustainability and ESG data \citep{lyon2015, oecd2020, asif2023} by proposing a framework linking specific AI technologies to potentially ovecome ESG data limitations.

The remainder of this paper is organized as follows. Section~\ref{sec:methodology} describes the methodology, including search strategy, inclusion criteria, and the resulting corpus, provides a structured taxonomy and detailed background on AI approaches applicable to sustainable finance, including an extensive review of the literature organized by methodology. Section~\ref{sec:discussion} discusses the results and outlines future research directions. Finally, Section~\ref{sec:conclusion} concludes the paper.

\section{Methodology}
\label{sec:methodology}


We offer an integrative or critical review on the intersection between sustainable finance and AI applications literatures. Due to the interdisciplinary nature of our inquiry, an integrative review is the most suitable methodology to respond to our research questions \citep{snyder2019}. Integrative reviews differ from systematic literature reviews in that the review coverage is representative as opposed to exhaustive \citep{cooper1988, torraco2016} ``to combine perspectives to create new theoretical models'' \citep[p.~334]{snyder2019}. Moreover, for emerging themes such as AI, integrative literature reviews can combine insights from different fields and create preliminary conceptualizations and frameworks \citep{snyder2019, torraco2016, webster2002}.

To review the literature on sustainable finance and AI, we followed a phase process \citep{snyder2019}. The first stage determines the inclusion criteria. For doing so, we delineate the boundaries of our analysis by contextualizing sustainable finance as the activities defined in the \citet{unfccc2015} framework for sustainable finance. Since there is not a standard definition for sustainable finance \citep{oecd2020} a limited keyword search would leave potential areas unexplored, thus we implement an overarching key word string based on \citet{singhania2025} coupled with AI systems keywords. Due to the emergence of the topic, we do not impose a temporal structure on our search \citep{torraco2016}. Our literature search\footnote{We use two queries to build the combined sample of articles:

Query 1: ("Blended financ*" OR "Blue financ*" OR "Carbon financ*" OR "Climate awareness bonds" OR "Climate financ*" OR "Sustainable digital financ*" OR "Ecological financ*" OR "Energy financ*" OR "Environmental financ*" OR "ESG financ*" OR "Ethical financ*" OR "Green Bank*" OR "Green bond" OR "Green financ*" OR "Impact financ*" OR "Private climate financ*" OR "Responsible financ*" OR "Social bonds" OR "social financ*" OR "Social Impact bonds" OR "Social impact financ*" OR "Socially responsible financ*" OR "sustainable bank*" OR "Sustainability bonds" OR "Sustainable corporate financ*" OR "Sustainable financ*" OR "Transition financ*" OR "Value based financ*" OR "Thematic bonds" OR "sovereign green bonds" OR "blue bonds" OR "climate bonds" OR "green loans" OR "SDG linked bonds" OR "sustainability linked loans" OR "sustainability linked bonds" OR "transition bonds" OR "social loans" OR "SDG financ*") AND (AB=("AI" OR "ML" OR "NLP" OR "Artificial intelligence" OR "Machine learning" OR "Natural language processing" OR "big data"))

Query 2: (ESG) AND (AB=("AI" OR "ML" OR "NLP" OR "LLM" OR "Artificial intelligence" OR "Machine learning" OR "Natural language processing" OR "big data" OR "Large Language Models"))} in the WoS database \citep{zupic2015} yielded 544 articles.

Studies were included in the review if they satisfied the following criteria: i) The work applies or proposes an AI, machine learning, NLP, or optimization methodology to an ESG-related problem in the financial domain; ii) The publication is a peer-reviewed journal article, and, iii) The work is written in English.

In turn, studies were excluded if they: i) Focus exclusively on traditional statistical or econometric methods without an AI component, ii) Address sustainability topics outside the financial domain (e.g., pure environmental science without financial application), or, iii) Are duplicates, editorials, or short opinion pieces without empirical or methodological contribution.

In the second phase we conduct the preliminary review by reading abstracts first and then full texts to verify that the articles meet the inclusion criteria and were consistent with the sustainable finance theme. We apply an ambition lens, focusing on how AI can foster sustainable finance, as opposed to a pragmatic approach interested in the use of AI as a research method -for example, to build a database using large language models-. The multidisciplinar authors' team proved particularly useful to discern the exclusion or retention of documents based on their conceptual and methodological pertinence. This left us with 394 papers published between 2018 and 2025. We applied a backward snowball technique search browsing the references in the retained publications, which added another 4 relevant publications. Finally, our sample consists in 398 papers, in 9 conference proceedings and in 61 journals; 15 of them in finance journals or proceedings, 19 in economy and/or management related areas journals and 36 in other disciplines.

In the third phase we use a conceptual or thematic structure to organize and code the review \citep{paul2020, paul2021spar} through content analysis. We select the Antecedents-Decisions-Outcomes (ADO) framework \citep{lim2021, paulbenito2018} to guide our review and identify the main themes emanating from the literature and their relationships. The final sample of \textbf{398 }papers were read in its entirety several times. We went back and forward between the data and the emerging concepts and perspectives until we found unifying themes that fit the ADO framework. As antecedents or reasons, we consider the different ESG grand challenges that require a sustainable finance solution, and their associated sub-domains//sub-dimensions. For example, data gathering through big data and sentiment analysis in social media or satellite imagery to assess a firm's carbon footprint. For the second ADO pillar, Decisions or execution, we encapsulate AI affordabilities such as data management, ESG risk assessment, scoring, portfolio optimization or sentiment analysis, among other. Finally, outcomes in the ADO framework were subdivided into outcomes for ESG analyses and for specific investments (such as portfolio management).

Each document was coded independently by two different authors. To ensure reliability, coders discussed their differences and recoded possible deviations \citep{snyder2019}. The outcome of the content analysis, the AI taxonomy and the emerging framework for AI powered sustainable finance is reported in the next sections.

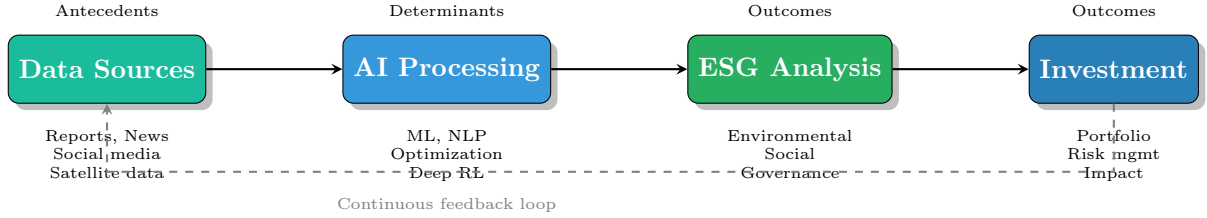
\begin{figure}[H]
\centering
\begin{tikzpicture}[
    node distance=1cm,
    box/.style={rectangle, rounded corners, draw, fill=#1, text=white, font=\footnotesize\bfseries, minimum width=2cm, minimum height=0.9cm, align=center, drop shadow},
    arrow/.style={->, thick, >=stealth}
]

\node[box=bigdatateal] (data) {Data Sources};
\node[below=0.2cm of data, font=\tiny, align=center, text width=2cm] {Reports, News\\Social media\\Satellite data};
\node[above=0.1cm of data, font=\tiny, align=center, text width=2cm] {Antecedents};

\node[box=aiblue, right=1.8cm of data] (ai) {AI Processing};
\node[below=0.2cm of ai, font=\tiny, align=center, text width=2cm] {ML, NLP\\Optimization\\Deep RL};
\node[above=0.1cm of ai, font=\tiny, align=center, text width=2cm] {Determinants};

\node[box=esggreen, right=1.8cm of ai] (esg) {ESG Analysis};
\node[below=0.2cm of esg, font=\tiny, align=center, text width=2cm] {Environmental\\Social\\Governance};
\node[above=0.1cm of esg, font=\tiny, align=center, text width=2cm] {Outcomes};

\node[box=finblue, right=1.8cm of esg] (invest) {Investment};
\node[below=0.2cm of invest, font=\tiny, align=center, text width=2cm] {Portfolio\\Risk mgmt\\Impact};
\node[above=0.1cm of invest, font=\tiny, align=center, text width=2cm] {Outcomes};

\draw[arrow] (data) -- (ai);
\draw[arrow] (ai) -- (esg);
\draw[arrow] (esg) -- (invest);

\draw[arrow, dashed, gray] (invest.south) -- ++(0,-0.9) -| (data.south);
\node[below=1.1cm of ai, font=\tiny, gray] {Continuous feedback loop};

\end{tikzpicture}
\caption{ADO framework of AI integration in ESG-driven sustainable finance. Data from multiple sources is processed by AI systems to generate ESG assessments that inform investment decisions, with continuous feedback loops for model improvement.}
\label{fig:overview}
\end{figure}

Figure~\ref{fig:overview} illustrates the Antecedents-Decisions-Outcomes (ADO) framework of AI integration in ESG-driven sustainable finance that guides our review.

\section{A Taxonomy on AI Applications to Sustainable Finance}
\label{sec:background}

This section provides a structured taxonomy of suitable AI applications to sustainable finance, categorizing them based on their underlying models and algorithms and their impact on sustainable investment decision-making processes. This taxonomy not only serves as a guide to the current state of AI in sustainable finance, but also highlights the future potential and challenges of these technologies.

\begin{figure}[H]
\centering
\resizebox{\textwidth}{!}{%
\begin{tikzpicture}[
    root/.style={rectangle, rounded corners, draw, fill=aiblue, text=white, font=\small\bfseries, minimum width=3cm, minimum height=0.7cm, align=center},
    l1/.style={rectangle, rounded corners, draw, fill=mlgreen, text=white, font=\footnotesize\bfseries, minimum width=2.2cm, minimum height=0.55cm, align=center},
    l2/.style={rectangle, rounded corners, draw, fill=dlpurple, text=white, font=\scriptsize\bfseries, minimum width=1.6cm, minimum height=0.5cm, align=center},
    l3/.style={rectangle, rounded corners, draw, fill=rlorange, text=white, font=\scriptsize, minimum width=1.5cm, minimum height=0.45cm, align=center},
    arrow/.style={->, thick, >=stealth}
]

\node[root] at (0,0) (ai) {AI Technologies};

\node[l1] at (-8,-1.5) (ml) {Machine Learning};
\node[l1] at (0,-1.5) (nlp) {NLP \& LLMs};
\node[l1] at (8,-1.5) (opt) {Optimization};

\draw[arrow] (ai.south) -- ++(0,-0.5) -| (ml.north);
\draw[arrow] (ai.south) -- (nlp.north);
\draw[arrow] (ai.south) -- ++(0,-0.5) -| (opt.north);

\node[l2] at (-10.5,-3) (sup) {Supervised};
\node[l2] at (-8,-3) (unsup) {Unsupervised};
\node[l2] at (-5.5,-3) (rl) {Reinforced};

\draw[arrow] (ml.south) -- ++(0,-0.5) -| (sup.north);
\draw[arrow] (ml.south) -- (unsup.north);
\draw[arrow] (ml.south) -- ++(0,-0.5) -| (rl.north);

\node[l3] at (-12.5,-4.5) (ens) {Ensembles};
\node[l3] at (-10.5,-4.5) (deep) {Deep Learning};
\node[l3] at (-8,-4.5) (clust) {Clustering};
\node[l3] at (-5.5,-4.5) (drl) {DQN/A2C};

\draw[arrow] (sup.south) -- ++(0,-0.5) -| (ens.north);
\draw[arrow] (sup.south) -- ++(0,-0.5) -| (deep.north);
\draw[arrow] (unsup.south) -- (clust.north);
\draw[arrow] (rl.south) -- (drl.north);

\node[l2] at (-2.8,-3) (trans) {Transformers};
\node[l2] at (0,-3) (sent) {Sentiment};
\node[l2] at (2.8,-3) (class) {Classification};

\draw[arrow] (nlp.south) -- ++(0,-0.5) -| (trans.north);
\draw[arrow] (nlp.south) -- (sent.north);
\draw[arrow] (nlp.south) -- ++(0,-0.5) -| (class.north);

\node[l3] at (-2.8,-4.5) (bert) {BERT/GPT};
\node[l3] at (0,-4.5) (opinion) {Opinion};
\node[l3] at (2.8,-4.5) (esgclass) {ESG Risks};

\draw[arrow] (trans.south) -- (bert.north);
\draw[arrow] (sent.south) -- (opinion.north);
\draw[arrow] (class.south) -- (esgclass.north);

\node[l2] at (6.5,-3) (bo) {Bayesian Optimization};
\node[l2] at (10,-3) (ga) {Evolutionary};

\draw[arrow] (opt.south) -- ++(0,-0.5) -| (bo.north);
\draw[arrow] (opt.south) -- ++(0,-0.5) -| (ga.north);

\node[l3] at (6.5,-4.5) (gp) {Gaussian Process};
\node[l3] at (10,-4.5) (nsga) {Genetic Algorithms};

\draw[arrow] (bo.south) -- (gp.north);
\draw[arrow] (ga.south) -- (nsga.north);

\end{tikzpicture}%
}
\caption{AI Technologies Taxonomy that are applied for ESG factors during these years. Classical Machine Learning paradigms including supervised, unsupervised, and reinforcement learning. We consider another category for Natural Language Processing and Large Language Models and another one regarding Optimization methods.}
\label{fig:ai_taxonomy}
\end{figure}
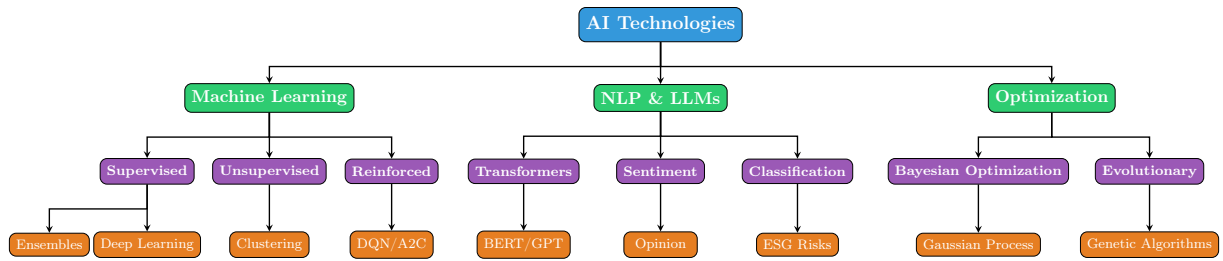

\begin{figure}[H]
\centering
\resizebox{\textwidth}{!}{%
\begin{tikzpicture}[
    root/.style={rectangle, rounded corners, draw, fill=bigdatateal, text=white, font=\small\bfseries, minimum width=3cm, minimum height=0.7cm, align=center},
    l2/.style={rectangle, rounded corners, draw, fill=dlpurple, text=white, font=\footnotesize\bfseries, minimum width=2.4cm, minimum height=0.55cm, align=center},
    l3/.style={rectangle, rounded corners, draw, fill=rlorange, text=white, font=\scriptsize, minimum width=2cm, minimum height=0.5cm, align=center},
    arrow/.style={->, thick, >=stealth}
]

\node[root] at (0,0) (bd) {Big Data Technologies};

\node[l2] at (-8,-1.5) (aug) {Data Augmentation};
\node[l2] at (0,-1.5) (int) {Integration of Sources};
\node[l2] at (7,-1.5) (qual) {Qualitative Studies};

\draw[arrow] (bd.south) -- ++(0,-0.5) -| (aug.north);
\draw[arrow] (bd.south) -- (int.north);
\draw[arrow] (bd.south) -- ++(0,-0.5) -| (qual.north);

\node[l3] at (-9.5,-3) (synth) {Synthetic Data};
\node[l3] at (-6.5,-3) (transfer) {Transfer Learning};

\draw[arrow] (aug.south) -- ++(0,-0.5) -| (synth.north);
\draw[arrow] (aug.south) -- ++(0,-0.5) -| (transfer.north);

\node[l3] at (-3.5,-3) (struct) {Structured Data};
\node[l3] at (-0.6,-3) (unstruct) {Unstructured Data};
\node[l3] at (2.5,-3) (alt) {Alternative Data};

\draw[arrow] (int.south) -- ++(0,-0.5) -| (struct.north);
\draw[arrow] (int.south) -- ++(0,-0.5) -| (unstruct.north);
\draw[arrow] (int.south) -- ++(0,-0.5) -| (alt.north);

\node[l3] at (5.5,-3) (expert) {Expert Interviews};
\node[l3] at (8.5,-3) (cases) {Case Studies};

\draw[arrow] (qual.south) -- ++(0,-0.5) -| (expert.north);
\draw[arrow] (qual.south) -- ++(0,-0.5) -| (cases.north);

\end{tikzpicture}%
}
\caption{Big Data Technologies Taxonomy. Data augmentation expands limited datasets through synthetic generation and transfer learning. Integration of sources combines structured financial data with unstructured text and alternative data. Qualitative studies incorporate expert knowledge and case-based insights.}
\label{fig:bigdata_taxonomy}
\end{figure}
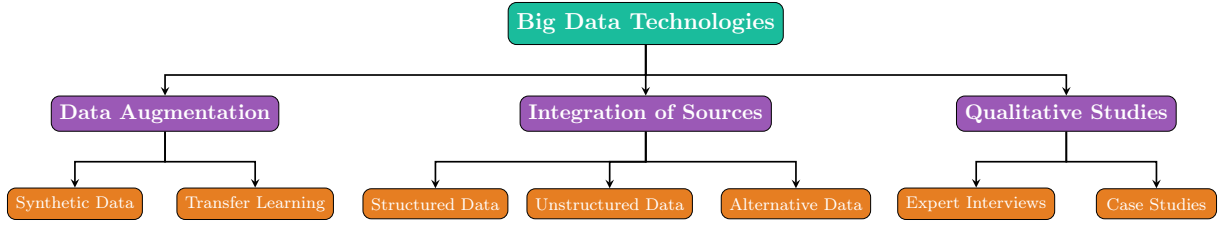

We present the taxonomy of AI applications in Figure~\ref{fig:ai_taxonomy}. This taxonomy provides a structured overview of AI technologies applicable in sustainable finance, highlighting the diverse methodologies that can be employed for enhanced sustainable investment decision-making. Each technology brings unique strengths to the table, collectively contributing to more informed, sustainable, and strategic financial investments.

AI is the field of computer science dedicated to creating systems capable of performing tasks that typically require human intelligence by simulating it. These tasks include learning, reasoning, problem-solving, generalizing data, and generating multi-modal data \citep{russell2010}. Within AI technologies, we first differentiate between proper AI modelling and the necessary big data technologies that are required to implement AI for ESG applications. Within AI modelling, we present machine learning main paradigms,  natural language processing and optimization, as first-level categories, which can in turn be decomposed into further sub-categories. Similarly, we divide big data into data augmentation, integration of sources, and qualitative studies as we illustrate in Figure \ref{fig:bigdata_taxonomy}.

Table~\ref{tab:literature} presents a comprehensive organization of the reviewed literature according to the categories of our proposed taxonomy, providing researchers with a structured reference for identifying relevant prior work in each methodological area.

\begin{table}[!htbp]
\centering
\caption{Organization of reviewed literature by AI technology and ESG application domain.}
\label{tab:literature}
\small
\begin{tabular}{p{3.5cm}p{4cm}p{6cm}}
\toprule
\textbf{AI Technology}& \textbf{ESG Application Domain}& \textbf{References} \\
\midrule
\multirow{5}{3.5cm}{Machine Learning. Supervised Learning (Ensemble Methods)}
    & Financial distress prediction & \citet{citterio2023} \\
    & ESG score prediction & \citet{damato2021} \\
    & Controversy prediction & \citet{lheureux2023} \\
    & Corporate governance rating & \citet{svanberg2022} \\
    & ESG-financial performance & \citet{delucia2020} \\
\midrule
\multirow{3}{3.5cm}{Machine Learning. Supervised Learning (Deep Learning)}
    & Stock return prediction with DL & \citet{meng2022} \\
    & CSR prediction & \citet{teoh2019} \\
\midrule
\multirow{2}{3.5cm}{Machine Learning. Unsupervised Learning}
    & ESG theme extraction & \citet{jaiswal2024} \\
    & Climate litigation clustering & \citet{raghupathi2023} \\
\midrule
\multirow{4}{3.5cm}{Machine Learning. Reinforcement Learning}
    & ESG portfolio management & \citet{garrido2023drl}, \citet{maree2022}, \citet{day2023} \\
    & Multi-objective Deep RL ESG optim. & \citet{garrido2024mobo} \\
    & Deep RL Explainable portfolio policies & \citet{delarica2025} \\
\midrule
\multirow{4}{3.5cm}{Natural Language Processing}
    & Sustainability report analysis & \citet{gupta2024} \\
    & ESG risk classification & \citet{kazakov2023} \\
    & ESG document categorization & \citet{lee2024} \\
    & Climate risk disclosure & \citet{garrido2023climatebert} \\
    & News sentiment analysis & \citet{pikatza2024} \\
\midrule
\multirow{2}{3.5cm}{Optimization}
    & Bayesian portfolio optim. & \citet{garrido2023bo} \\
    & Genetic algorithm optim. & \citet{garcia2022} \\
\bottomrule
\end{tabular}
\end{table}

\subsection{Machine Learning Technologies for ESG }

Machine learning \citep{james2023} encompasses algorithms that learn from and make predictions or decisions based on data without being explicitly programmed. These models can be used to solve tasks like regression, classification, clustering, policy estimations, and planning under uncertainty. Machine learning encompasses three fundamental paradigms: supervised learning, where models learn from labeled datasets; unsupervised learning, where models discover patterns in unlabeled data; and reinforcement learning, where agents learn optimal policies through interaction with an environment and reward signals.

Supervised learning models learn to solve regression and classification tasks from labeled data, helping to predict outcomes or classify data into categories \citep{bishop2006}. Within this paradigm, ensemble methods have demonstrated particularly strong performance in ESG applications due to their ability to handle tabular financial data effectively. \citet{citterio2023} conducted a comprehensive study applying random forest models to forecast bank financial distress using ESG data, demonstrating superior predictive accuracy relative to traditional financial ratios and logistic regression, with clear implications for financial stability monitoring.

\citet{guptayan2025} analyze the relationship between ESG-related risks and firm volatility through a machine learning framework that integrates ESG, geopolitical, and supply chain disruption risks. These three emerging sources of risk are extracted from corporate disclosures using large language models (LLMs), allowing the transformation of unstructured textual information into quantifiable indicators. The study combines these novel risk measures with traditional financial variables within an LSTM neural network to predict firm-specific volatility. The integration of ESG, geopolitical, and supply chain risks is particularly relevant in the current environment, where firms are increasingly exposed to interconnected non-financial risks that may significantly affect financial stability and risk dynamics.

The challenge of missing ESG scores represents a significant obstacle for sustainable finance research and practice, as many companies, particularly small and medium enterprises and those in emerging markets, lack comprehensive coverage from ESG rating agencies. \citet{damato2021} addressed this gap by developing machine learning models that predict ESG scores using fundamental financial ratios as inputs, enabling analysts to estimate sustainability performance for companies without official ratings. Their approach leverages the correlation between financial characteristics and ESG behavior, providing a practical solution for investors seeking to expand their sustainable investment universe beyond the limited set of rated companies. Building on this line of work, \citet{martinmelero2025} compared the predictive value of sectorial versus financial data sources for ESG scoring of mutual funds, finding that combining both data types within machine learning models yields more reliable ESG estimates than either source alone.

ESG controversies, such as environmental scandals, labor disputes, or governance failures, can have devastating effects on company valuations and investor portfolios. \citet{lheureux2023} developed a sophisticated system for predicting ESG controversies by analyzing Twitter sentiment using machine learning techniques. By monitoring social media discussions and detecting early warning signals, their approach enables investors to anticipate potential controversies before they fully materialize in traditional news sources or rating agency assessments. This real-time monitoring capability represents a significant advancement over static ESG ratings that are typically updated only quarterly or annually.

Corporate greenwashing is a key issue regarding sustainable corporate behavior. \citet{song2025} analyze it in Chinese heavy-pollution industries and find that predictive models can improve the identification of deceptive environmental practices by more than 20\% compared with benchmark approaches. By combining external, organizational, and managerial firm characteristics, the analysis seeks to forecast the extent of greenwashing and provide a more effective tool for evaluating corporate environmental claims. The results highlight the potential of data-driven approaches to enhance transparency and support sustainable corporate behavior. To achieve this, the study employs several machine learning techniques, including random forest, support vector machines, K-nearest neighbors, artificial neural networks, and linear regression models.

Corporate governance performance has traditionally been assessed through qualitative analysis by specialized rating agencies, but \citet{svanberg2022} demonstrated that machine learning can automate and potentially improve this process. Using linear models and more complex algorithms, they developed predictive systems for corporate governance ratings that achieve high accuracy while providing interpretable explanations of the factors driving governance assessments. This work contributes to the ongoing effort to standardize and objectify ESG measurement across the financial industry.

The question of whether strong ESG performance leads to superior financial returns remains one of the most debated topics in sustainable finance. \citet{delucia2020} employed Support Vector Machines and k-nearest neighbors algorithms to investigate this relationship empirically, analyzing public enterprises across Europe. Their machine learning approach enables nonlinear modeling of the ESG-performance relationship, revealing nuanced patterns that linear regression would miss. The findings provide evidence that ESG factors do contribute to financial outperformance, though the relationship varies across sectors and time periods.

Deep learning, a subset of machine learning involving neural networks with many layers \citep{goodfellow2016, lecun2015}, has revolutionized many fields and is increasingly being applied to sustainable finance challenges. Deep learning can integrate large amounts of structured, unstructured, and multimodal data such as video, audio, or satellite imagery, making it particularly suited for the heterogeneous data landscape of ESG analysis. \citet{meng2022} applied deep learning architectures for stock excess return prediction incorporating ESG factors, demonstrating that neural networks can capture complex nonlinear relationships between sustainability metrics and financial performance that simpler models would overlook. Their model integrates ESG data alongside traditional financial features, providing evidence that sustainability information contains alpha-generating signals for equity investors.

\citet{teoh2019} explored the application of deep learning for corporate social responsibility prediction, developing models that can forecast future CSR performance based on historical patterns and company characteristics. This predictive capability is valuable for investors seeking to identify companies likely to improve their sustainability profiles, enabling forward-looking rather than backward-looking ESG investment strategies. Beyond return prediction, deep learning has also been applied to the estimation of realized covariance matrices, a critical input for risk management and portfolio construction. \citet{souto2026} proposed a multivariate neural network architecture that improves covariance matrix forecasts relative to standard econometric benchmarks, with direct implications for the risk modelling stage that precedes ESG-aware portfolio allocation.

Unsupervised learning involves training AI systems on data without labeled responses \citep{hastie2009}, enabling the discovery of hidden patterns and structures. This approach is particularly valuable in ESG contexts where predefined categories may not capture the full complexity of sustainability issues. \citet{jaiswal2024} applied unsupervised learning techniques to extract the main ESG themes being discussed on Twitter, revealing the key topics that concern investors and the public regarding corporate sustainability. Their analysis identified distinct clusters of ESG discourse, providing insights into how different stakeholder groups conceptualize and prioritize sustainability issues. This understanding of public perception is crucial for companies seeking to align their ESG communications with stakeholder expectations.

Climate change litigation has emerged as a significant risk factor for companies in carbon-intensive industries, and understanding the patterns in legal cases is essential for risk assessment. \citet{raghupathi2023} employed clustering algorithms to analyze climate change-related litigation cases from the Westlaw database, identifying four main clusters of cases based on their characteristics and legal arguments. This systematic categorization helps legal analysts and risk managers understand the landscape of climate litigation and anticipate potential legal exposures for their portfolio companies. Beyond physical and litigation risk, machine learning is also being deployed to prioritise risk management strategies in sustainability-aligned projects. \citet{eti2026} combined machine learning with a fuzzy decision-making framework to rank financial risk management techniques for renewable energy investments, illustrating how hybrid AI approaches can support project-level risk assessment in clean-energy finance.

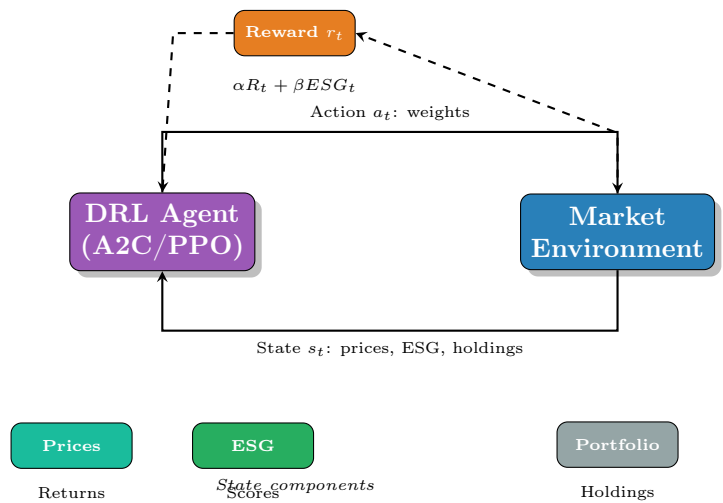
\begin{figure}[H]
\centering
\begin{tikzpicture}[
    node distance=1cm,
    box/.style={rectangle, rounded corners, draw, fill=#1, text=white, font=\footnotesize\bfseries, minimum width=2.4cm, minimum height=1cm, align=center, drop shadow},
    smallbox/.style={rectangle, rounded corners, draw, fill=#1, text=white, font=\tiny\bfseries, minimum width=1.6cm, minimum height=0.6cm, align=center},
    arrow/.style={->, thick, >=stealth}
]

\node[box=dlpurple] (agent) {DRL Agent\\(A2C/PPO)};

\node[box=finblue, right=3.5cm of agent] (env) {Market\\Environment};

\draw[arrow] (agent.north) -- ++(0,0.8) -| node[pos=0.25, above, font=\tiny] {Action $a_t$: weights} (env.north);

\draw[arrow] (env.south) -- ++(0,-0.8) -| node[pos=0.25, below, font=\tiny] {State $s_t$: prices, ESG, holdings} (agent.south);

\node[smallbox=rlorange, above=1.8cm of agent, xshift=1.75cm] (reward) {Reward $r_t$};
\node[below=0.15cm of reward, font=\tiny, align=center] {$\alpha R_t + \beta ESG_t$};

\draw[arrow, dashed] (env.north) -- ++(0,0.8) -- (reward.east);
\draw[arrow, dashed] (reward.west) -- ++(-0.8,0) -- (agent.north);

\node[smallbox=bigdatateal, below=2cm of agent, xshift=-1.2cm] (prices) {Prices};
\node[smallbox=esggreen, below=2cm of agent, xshift=1.2cm] (esg) {ESG};
\node[smallbox=optgray, below=2cm of env] (portfolio) {Portfolio};

\node[below=0.1cm of prices, font=\tiny] {Returns};
\node[below=0.1cm of esg, font=\tiny] {Scores};
\node[below=0.1cm of portfolio, font=\tiny] {Holdings};

\node[below=2.6cm of agent, xshift=1.75cm, font=\tiny\itshape] {State components};

\end{tikzpicture}
\caption{Deep Reinforcement Learning framework for ESG-integrated portfolio management. The agent observes market states including ESG scores, takes actions (portfolio rebalancing), and receives rewards that combine financial returns with ESG performance.}
\label{fig:drl}
\end{figure}

Reinforcement learning represents one of the most promising frontiers for AI in sustainable finance, particularly for portfolio management applications. In this paradigm, an autonomous agent learns to make decisions by trial and error, performing actions in an environment to achieve cumulative rewards that guide policy development \citep{sutton2018}. Deep reinforcement learning, which combines reinforcement learning with deep neural networks \citep{mnih2015}, has enabled agents to learn complex policies directly from high-dimensional inputs. Unlike traditional portfolio optimization approaches such as the Markowitz mean-variance model \citep{markowitz1952}, reinforcement learning makes no assumptions about the distribution of returns and can incorporate arbitrary information sources including ESG scores. Figure~\ref{fig:drl} illustrates the deep reinforcement learning framework for ESG portfolio management.

\citet{garrido2023drl} conducted pioneering research on Deep Reinforcement Learning for ESG financial portfolio management, investigating the potential benefits of ESG score-based market regulation. Using an Advantage Actor-Critic agent, one of the policy gradient methods alongside Proximal Policy Optimization \citep{schulman2017}, within environments adapted from the FinRL platform \citep{liu2020finrl} on the OpenAI Gym framework, they performed experiments comparing DRL agent performance under standard Dow Jones Industrial Average market conditions against a scenario where returns are regulated according to company ESG scores. In the ESG-regulated market simulation, grants were proportionally allocated to portfolios based on their returns and ESG scores, while taxes were assigned to portfolios below the mean ESG score of the index. The results demonstrated that the DRL agent operating within the ESG-regulated market outperformed the standard market setup, suggesting that ESG-aware policies can achieve superior risk-adjusted returns. This finding has profound implications for policymakers considering regulatory interventions to promote sustainable investment.

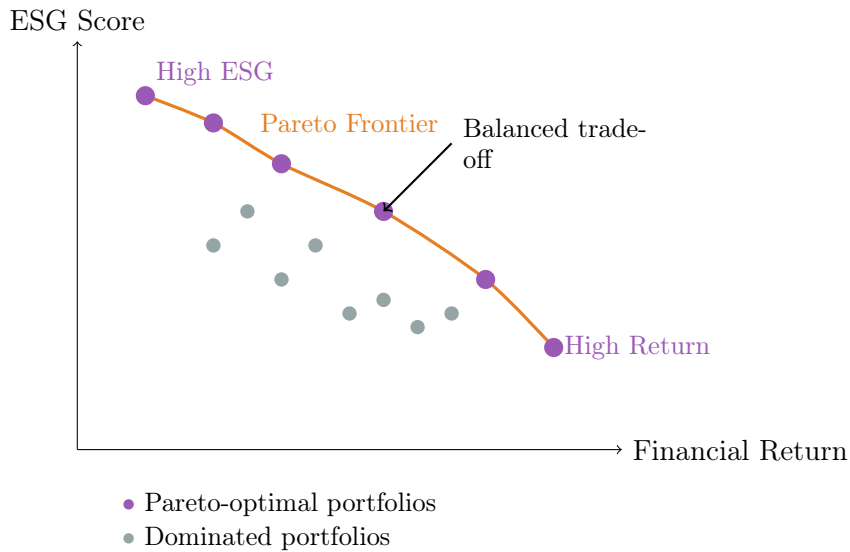
\begin{figure}[H]
\centering
\begin{tikzpicture}[scale=0.9]
    \draw[->] (0,0) -- (8,0) node[right, font=\small] {Financial Return};
    \draw[->] (0,0) -- (0,6) node[above, font=\small] {ESG Score};

    \draw[very thick, rlorange, smooth] plot coordinates {(1,5.2) (2,4.8) (3,4.2) (4.5,3.5) (6,2.5) (7,1.5)};

    \fill[dlpurple] (1,5.2) circle (4pt);
    \fill[dlpurple] (2,4.8) circle (4pt);
    \fill[dlpurple] (3,4.2) circle (4pt);
    \fill[dlpurple] (4.5,3.5) circle (4pt);
    \fill[dlpurple] (6,2.5) circle (4pt);
    \fill[dlpurple] (7,1.5) circle (4pt);

    \fill[optgray] (2,3) circle (3pt);
    \fill[optgray] (3,2.5) circle (3pt);
    \fill[optgray] (4,2) circle (3pt);
    \fill[optgray] (5,1.8) circle (3pt);
    \fill[optgray] (3.5,3) circle (3pt);
    \fill[optgray] (4.5,2.2) circle (3pt);
    \fill[optgray] (2.5,3.5) circle (3pt);
    \fill[optgray] (5.5,2) circle (3pt);

    \node[above right, font=\footnotesize, dlpurple] at (1,5.2) {High ESG};
    \node[right, font=\footnotesize, dlpurple] at (7,1.5) {High Return};
    \node[above, font=\footnotesize, rlorange] at (4,4.5) {Pareto Frontier};

    \node[right, font=\footnotesize] at (0.5,-0.8) {\textcolor{dlpurple}{$\bullet$} Pareto-optimal portfolios};
    \node[right, font=\footnotesize] at (0.5,-1.3) {\textcolor{optgray}{$\bullet$} Dominated portfolios};

    \draw[<-, thick] (4.5,3.5) -- (5.5,4.5) node[right, font=\footnotesize, text width=2.5cm] {Balanced trade-off};

\end{tikzpicture}
\caption{Multi-objective optimization for ESG portfolio management. The Pareto frontier represents portfolios where improving one objective (financial return or ESG score) necessarily worsens the other. Bayesian optimization efficiently identifies these optimal trade-off solutions.}
\label{fig:pareto}
\end{figure}

Building upon this foundation, \citet{garrido2024mobo} developed a multi-objective Bayesian optimization framework for tuning Deep Reinforcement Learning agents in ESG portfolio management contexts. The challenge with DRL agents is that their performance exhibits high variability and extreme sensitivity to hyperparameter values. Training a single agent requires millions of timesteps, making exhaustive hyperparameter search computationally prohibitive. Their approach treats hyperparameter optimization as a multi-objective problem, simultaneously maximizing portfolio return, minimizing risk, and maximizing ESG score. Using Bayesian optimization \citep{shahriari2016}, which is particularly suited for expensive black-box function optimization, they efficiently explore the hyperparameter space to identify Pareto-optimal configurations that balance these competing objectives. This methodology enables practitioners to select DRL agents according to their specific preferences for the trade-offs between financial performance and sustainability impact. Figure~\ref{fig:pareto} illustrates the concept of multi-objective optimization in ESG portfolio management.

The interpretability of AI-driven investment decisions has become increasingly important as regulators and clients demand transparency. \citet{delarica2025} addressed this challenge by developing explainable post-hoc analysis methods for Deep Reinforcement Learning portfolio management policies. Their work acknowledges that while DRL approaches can successfully manage portfolios in high-volatility markets where traditional Markowitz assumptions \citep{markowitz1952} break down, the resulting policies are often opaque. By applying explanation techniques such as SHAP \citep{lundberg2017} and LIME \citep{ribeiro2016} to trained DRL agents, they enable human understanding of why the agent makes specific trading decisions, which ESG factors influence portfolio construction, and how the policy adapts to changing market conditions. This explainability is essential for institutional adoption of AI-driven sustainable investment strategies.

\citet{maree2022} explored the balance between profit, risk, and sustainability in reinforcement learning portfolio management, developing agents that explicitly incorporate multiple objectives in their reward functions. Their research demonstrates that reinforcement learning can effectively navigate the trade-offs between financial returns and ESG impact, finding portfolio allocations that satisfy both profit-seeking and sustainability-conscious investors. \citet{day2023} extended this line of research by developing dynamic trading strategies using deep reinforcement learning that incorporate ESG scores, enabling portfolios to adapt to changing sustainability information over time.

\subsection{Natural language Technologies for ESG}

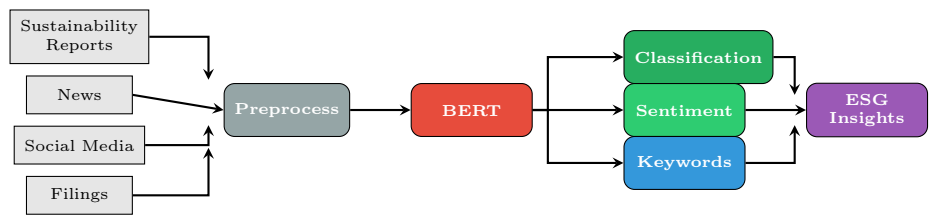
\begin{figure}[H]
\centering
\begin{tikzpicture}[
    node distance=0.6cm,
    box/.style={rectangle, rounded corners, draw, fill=#1, text=white, font=\tiny\bfseries, minimum width=1.6cm, minimum height=0.7cm, align=center},
    doc/.style={rectangle, draw, fill=gray!20, font=\tiny, minimum width=1.4cm, minimum height=0.5cm, align=center},
    arrow/.style={->, thick, >=stealth}
]

\node[doc] (doc1) {Sustainability\\Reports};
\node[doc, below=0.15cm of doc1] (doc2) {News};
\node[doc, below=0.15cm of doc2] (doc3) {Social Media};
\node[doc, below=0.15cm of doc3] (doc4) {Filings};

\node[box=optgray, right=1.2cm of doc2, yshift=-0.2cm] (preproc) {Preprocess};

\node[box=nlpred, right=0.8cm of preproc] (bert) {BERT};

\node[box=esggreen, right=1.2cm of bert, yshift=0.7cm] (class) {Classification};
\node[box=mlgreen, right=1.2cm of bert] (sent) {Sentiment};
\node[box=aiblue, right=1.2cm of bert, yshift=-0.7cm] (key) {Keywords};

\node[box=dlpurple, right=0.8cm of sent] (output) {ESG\\Insights};

\draw[arrow] (doc1.east) -| ($(preproc.west)+(-0.2,0.4)$);
\draw[arrow] (doc2.east) -- (preproc.west);
\draw[arrow] (doc3.east) -| ($(preproc.west)+(-0.2,-0.2)$);
\draw[arrow] (doc4.east) -| ($(preproc.west)+(-0.2,-0.5)$);

\draw[arrow] (preproc) -- (bert);

\draw[arrow] (bert.east) -- ++(0.2,0) |- (class.west);
\draw[arrow] (bert.east) -- (sent.west);
\draw[arrow] (bert.east) -- ++(0.2,0) |- (key.west);

\draw[arrow] (class.east) -| ($(output.west)+(-0.15,0.2)$);
\draw[arrow] (sent.east) -- (output.west);
\draw[arrow] (key.east) -| ($(output.west)+(-0.15,-0.2)$);

\end{tikzpicture}
\caption{Natural Language Processing pipeline for ESG document analysis. Multiple text sources are preprocessed and analyzed using transformer models to extract ESG classifications, sentiment, and key themes.}
\label{fig:nlp_pipeline}
\end{figure}

Natural language processing has transformed the analysis of ESG-relevant textual information, which is abundant but historically difficult to process at scale. Large Language Models \citep{gozalo2023}, exemplified by systems like ChatGPT, are advanced deep learning models based on the transformer architecture \citep{vaswani2017} that have been trained on vast text corpora. These models can process and generate human-like text, making them invaluable for analyzing sustainability reports, news articles, regulatory filings, and social media discussions. Figure~\ref{fig:nlp_pipeline} presents the NLP pipeline for ESG document analysis.

\citet{gupta2024} applied the BERT transformer architecture \citep{devlin2019} combined with the YAKE keyword extraction technique for analyzing sustainability reports, enabling automated identification of the key themes and commitments disclosed by companies. Given that sustainability reports can span hundreds of pages and companies publish thousands of such documents annually, automated analysis is essential for systematic ESG assessment. Their approach extracts the most salient sustainability keywords and phrases, enabling analysts to quickly understand a company's ESG focus areas and track changes over time.

The task of classifying ESG risks from textual sources has been addressed by \citet{kazakov2023}, who developed ESGify, an automated system for categorizing environmental, social, and corporate governance risks from news and corporate disclosures. Their model can process large volumes of text and assign appropriate risk categories, enabling continuous monitoring of ESG exposures across portfolios. This automated classification replaces labor-intensive manual analysis while maintaining high accuracy.

\citet{lee2024} developed ESG2PreEM, an automated ESG grade assessment framework using pre-trained ensemble models. Their approach employs voting ensembles of different BERT transformers to categorize preprocessed ESG documents, improving upon the assessment criteria established by traditional rating agencies. By combining multiple transformer models, they achieve robust performance across diverse document types and ESG topics, demonstrating the potential for AI to enhance and potentially standardize ESG assessment processes.

Climate-related financial risk disclosure has become a regulatory priority following the recommendations of the Task Force on Climate-related Financial Disclosures \citep{tcfd2017}. \citet{garrido2023climatebert} addressed the challenge of analyzing climate risk disclosures by fine-tuning the ClimateBert transformer model \citep{webersinke2022} with the ClimaText dataset. Their specialized model can identify and categorize climate-related statements in corporate filings, enabling systematic assessment of how companies communicate their exposure to physical and transition risks. This capability is essential for investors seeking to understand climate risk across their portfolios and for regulators monitoring disclosure compliance.

Investors' sentiment towards corporate ESG disclosures is a main issue to understand due to the implications investors' sentiment has on stock prices. \citet{liao2025} provides an accurate insight in Taiwan's capital market, particularly emphasizing whether disclosure frequency or disclosure sentiment exerts a stronger influence on market perceptions. The analysis is conducted through an event study approach combined with a BERT-based natural language processing model. The findings show that investor reactions are more sensitive to the frequency of ESG communication than to sentiment. However, the effect varies across ESG dimensions.

\citet{pikatza2024} investigated the relationship between news content and ESG perceptions by analyzing which terms in news articles improve or worsen sympathy toward companies from an ESG perspective. Using natural language processing techniques, they identified linguistic patterns associated with positive and negative ESG sentiment, providing insights into how media coverage shapes public perception of corporate sustainability. This understanding enables companies to communicate more effectively about their ESG initiatives and helps investors contextualize news-driven ESG sentiment.

NLP techniques are also applied to analyze climate-related communication by central banks \citep{campiglio2025}. They show that stronger engagement with climate and green finance narratives is associated with higher equity returns for green firms relative to polluting firms. The findings also suggest that communication strategies are mainly shaped by institutional factors.

\subsection{Optimization AI technologies for ESG}

Optimization algorithms complement machine learning approaches by finding the best solutions within defined constraints, a capability essential for portfolio construction where investors must balance multiple objectives and satisfy regulatory or policy requirements. \citet{garrido2023bo} pioneered the application of Bayesian optimization to ESG portfolio investment, treating the portfolio as a computationally expensive black-box objective function. Traditional optimization approaches struggle with the complexity and noise inherent in ESG-integrated portfolios, but Bayesian optimization is specifically designed for such challenging scenarios. By building probabilistic surrogate models of the portfolio's performance as a function of allocation decisions, their approach efficiently searches for configurations that maximize risk-adjusted returns while satisfying ESG constraints. This work opened a significant new research direction in portfolio allocation methodology.

Evolutionary computation provides an alternative optimization paradigm that has proven effective for ESG portfolio construction. \citet{garcia2022} employed genetic algorithms for maximizing ESG performance in portfolio selection strategies applied to Dow Jones index stocks. Genetic algorithms maintain populations of candidate solutions that evolve through selection, crossover, and mutation operations, enabling exploration of complex solution spaces without gradient information. Their research quantified the cost of maximizing ESG performance in terms of financial return sacrifice, providing investors with explicit trade-off information for sustainable investment decisions.

The effective implementation of AI for ESG applications requires robust big data infrastructure and methodologies beyond the core algorithms. Data augmentation techniques can expand limited ESG datasets through synthetic data generation or transfer learning from related domains. Integration of heterogeneous sources---combining structured financial data with unstructured text, satellite imagery, social media, and IoT sensor data---enables more comprehensive sustainability assessment. Qualitative information from expert interviews, case studies, and stakeholder consultations can be incorporated into quantitative AI models through appropriate encoding schemes. These supporting technologies are essential for realizing the full potential of AI in sustainable finance.

\subsection{A framework on AI-powered sustainable finance: AI approaches addressing ESG data limitations and sustainable investment decisions}

The deployment of AI in sustainable finance is constrained by a set of structural limitations inherent to ESG data. Despite the rapid growth in ESG-related disclosures, prior research shows that sustainability data remains difficult to use due to persistent limitations related to coverage, credibility, uncertainty\textbf{, }structure\textbf{, }parsimony, and consistency, which collectively hinder comparability, reliability, and decision relevance \citep{chatterji2016, berg2022aggregate, lyon2015}. Table \ref{tab:placeholder} synthesizes these limitations by organizing them into a framework that shows a set of core ESG data attributes that constitute key barriers for AI-enabled sustainable finance applications.

\begin{table}[!htbp]
    \centering
    \scriptsize
    \setlength{\tabcolsep}{2pt}
    \renewcommand{\arraystretch}{1.1}
    \begin{tabular}{>{\raggedright\arraybackslash}p{2.2cm} >{\raggedright\arraybackslash}p{4.4cm} >{\raggedright\arraybackslash}p{1.9cm} >{\raggedright\arraybackslash}p{2.1cm} >{\raggedright\arraybackslash}p{1.9cm} >{\raggedright\arraybackslash}p{1.9cm} >{\raggedright\arraybackslash}p{1.6cm}}
    \toprule
         \textbf{Attribute} &  \textbf{Limitation} &  \textbf{ESG refs.} &  \textbf{Relevance} &  \textbf{AI solutions} &  \textbf{AI category} & \textbf{AI refs.} \\
    \midrule
         Coverage &  \textbf{Uncommensurability.} Incomplete and uneven data across firms, regions, and supply chains (e.g. Scope 3), leading to gaps. &  \citet{berg2022aggregate}; \citet{chatterji2016}; \citet{seles2018} &  Limits ESG risk assessment and portfolio impact analysis. &  Geospatial AI; data augmentation; multi-source integration &  ML (supervised/ unsupervised); optimization & \citet{seles2018} \\
         Credibility &  \textbf{Bias.} Selectively disclosed, self-reported, or unverified ESG info, including greenwashing. &  \citet{lyon2015}; \citet{chatterji2009} &  Undermines investor trust and ESG product integrity. &  NLP disclosure analysis; cross-source validation &  NLP; ML & \citet{burnaev2023} \\
         Uncertainty &  \textbf{Non-observability.} ESG risks depend on future developments (e.g. climate transition) not observable ex ante. &  \citet{bolton2021}; \citet{pankratz2023} &  Constrains forward-looking pricing and long-term allocation. &  Scenario modelling; stress testing; synthetic data &  ML; optimization & \citet{bang2023}; \citet{jabeur2021} \\
         Structure &  \textbf{Dispersion.} ESG info scattered across reports, filings, press releases, digital media. &  \citet{loughran2016}; \citet{kolbel2020} &  Increases processing costs, limits systematic integration. &  Text mining; information extraction &  NLP & \citet{burnaev2023} \\
         Parsimony &  \textbf{Fragmentation.} Proliferation of ESG standards, ratings, indices creates overload and complexity. &  \citet{camilleri2015}; \citet{berg2022aggregate} &  Reduces usability, increases costs for investors and firms. &  Dimensionality reduction; feature selection &  ML (representation learning) & \citet{agosto2023} \\
         Consistency &  \textbf{Heterogeneity.} ESG concepts and scores diverge across frameworks and rating providers. &  \citet{berg2022aggregate}; \citet{chatterji2016} &  Generates label noise and instability in ESG models. &  Ensemble learning; robustness analysis &  ML (ensemble methods) & \citet{agosto2023} \\
    \bottomrule
    \end{tabular}
    \caption{A framework on AI-powered sustainable finance. ESG data attributes, limitations, and AI solutions.}
    \label{tab:placeholder}
\end{table}

Building on the AI applications for sustainable finance framework presented above, Table \ref{tab:placeholder} also highlights  AI technologies that can help mitigate these ESG data challenges. In the following subsection, we present a set of illustrative examples of AI applications that address ESG data limitations in practice and support more informed sustainable finance decisions.

 The literature provides a growing number of examples illustrating how AI and advanced data-driven approaches can help mitigate the ESG data limitations summarized in Table \ref{tab:placeholder}. These contributions demonstrate how AI technologies improve the usability, comparability, and decision relevance of sustainability information by addressing challenges related to coverage, uncertainty, structure, parsimony, consistency, and credibility.

Several studies focus on ESG data limitations related to uncertainty, reflecting the forward-looking nature of many sustainability risks. \citet{bang2023} analyse ESG controversies and investor trading behaviour, showing how sustainability-related events introduce uncertainty that is not captured by static ESG scores. Their results illustrate how data-driven models can extract market-relevant ESG signals from event-based information, thereby improving the interpretation of uncertain and time-varying sustainability risks. Similarly, \citet{jabeur2021} examine the impact of environmental and energy-related factors on financial markets, highlighting how quantitative modelling frameworks can incorporate uncertain climate and environmental risks into financial analysis and investment decisions.

\citet{malik2025} analyze how country-level corruption influences the relationship between ESG performance and firm value. Using double-debiased machine learning and linear regression techniques, the authors find that stronger ESG performance is associated with higher market valuation, particularly in countries with lower corruption levels. The results highlight the importance of institutional quality in strengthening the value relevance of ESG practices.

Other contributions address limitations related to structure and the dispersion of ESG information across unstructured sources. \citet{burnaev2023} provide practical examples of AI applications for ESG challenges, emphasizing how machine learning and natural language processing can extract structured insights from textual sustainability disclosures, reports, and heterogeneous data sources. By reducing data dispersion and transforming unstructured ESG information into analyzable formats, these approaches support scalable ESG assessment and integration in sustainable finance.

Challenges associated with parsimony and informational overload are addressed in studies that apply probabilistic and machine learning models to high-dimensional ESG data. \citet{agosto2025} propose Bayesian learning models to measure the relationship between sustainability indicators and financial outcomes. Their approach illustrates how AI can reduce complexity by identifying and weighting the most decision-relevant ESG variables, thereby mitigating the fragmentation caused by the proliferation of ESG metrics, standards, and ratings. Such parsimonious representations enhance the interpretability and usability of ESG information for investors.

Limitations related to coverage and consistency also emerge across the literature. Big data and AI-enabled integration techniques discussed in \citet{seles2018} show how combining heterogeneous environmental data sources can reduce gaps and uncommensurability in sustainability information, particularly for complex issues such as supply-chain emissions. More broadly, AI-based modelling frameworks that integrate ESG indicators with financial data help reduce instability arising from heterogeneous ESG measures and divergent sustainability signals, improving consistency and robustness in sustainable finance applications.

Finally, credibility concerns related to selective disclosure and greenwashing have been widely documented in the literature \citep{lyon2015}. Text-based analytics and cross-source validation approaches enable the detection of inconsistencies between reported sustainability claims and underlying corporate behaviour by systematically analysing unstructured ESG disclosures \citep{loughran2016, kolbel2020, burnaev2023}, thereby strengthening confidence in ESG data used for financial decision-making.

\section{Discussion}
\label{sec:discussion}
This paper examines how artificial intelligence is being applied to sustainable finance and what role different AI technologies play in supporting ESG-driven investment decisions. Through an integrative review of the literature, the study systematically maps existing AI applications to sustainable finance objectives and identifies the main ESG data limitations these applications seek to address. The results show that AI adoption in sustainable finance is no longer exploratory but increasingly targeted at specific decision problems, particularly ESG assessment, risk prediction, portfolio construction, and monitoring.

A first key result is that AI applications in sustainable finance cluster clearly around three technological paradigms—machine learning, natural language processing, and optimization—each serving distinct but complementary functions. Machine learning approaches dominate applications related to prediction and scoring, including financial distress, ESG performance, and controversy detection, confirming their suitability for handling structured and tabular ESG–financial data \citep{citterio2023, damato2021, delucia2020}. Natural language processing methods are primarily employed to process unstructured sustainability disclosures, news, and social media, addressing the growing importance of textual ESG information and the limitations of traditional disclosure-based metrics \citep{loughran2016, kolbel2020, bingler2022}. Optimization and reinforcement learning approaches, in contrast, are mainly used at the decision stage, where trade-offs between financial returns and ESG objectives must be explicitly managed under uncertainty \citep{garrido2023bo, garrido2023drl, maree2022}.

A second important finding concerns the role of AI in mitigating ESG data limitations. Across the reviewed studies, AI is consistently used to cope with shortcomings in ESG data related to coverage, structure, parsimony, consistency, credibility, and uncertainty \citep{chatterji2016, berg2022aggregate}. Rather than eliminating these limitations, AI techniques tend to reduce their impact on decision-making by improving data integration, extracting signals from unstructured sources, and combining heterogeneous ESG indicators with financial information. For example, text-based models enhance the interpretability and credibility of ESG narratives by identifying inconsistencies and potential greenwashing \citep{lyon2015, burnaev2023}, while ensemble and probabilistic learning approaches mitigate instability arising from divergent ESG ratings \citep{agosto2023}.

\subsection{Future Research Directions}
\label{sec:future}

Based on our comprehensive review of the literature and analysis of current limitations, we identify several promising avenues for future research that could significantly advance the field of AI for sustainable finance.

As previously discussed, most AI applications have focused on analysing ESG-related issues in a broad sense. When examining separately each of the three components that constitute ESG—Environmental, Social, and Governance—it becomes evident that the Environmental dimension has been virtually the only one to receive significant attention and improvement through AI-based applications. By contrast, Governance and the Social pillar have remained largely unexplored. This constitutes a research area with substantial potential to benefit from the application of advanced techniques such as natural language processing (NLP), large language models (LLMs), and machine learning–based text analytics, which could generate meaningful improvements both in corporate management practices and in the development of indicators that may support regulatory monitoring and supervision.

Deep reinforcement learning represents one of the most promising yet still developing areas in ESG-integrated portfolio management, as demonstrated by the pioneering work of \citet{garrido2023drl}, \citet{garrido2024mobo}, and \citet{delarica2025}. Future research should focus on developing more sophisticated multi-objective reinforcement learning frameworks that explicitly balance financial returns with multiple ESG dimensions, moving beyond aggregate ESG scores to consider environmental, social, and governance factors separately. The creation of realistic simulation environments that accurately capture ESG-related market dynamics, including the impact of ESG controversies, regulatory changes, and shifting investor preferences, would enable more robust agent training. Transfer learning approaches that allow agents trained in one market or time period to adapt quickly to new conditions could reduce the enormous sample complexity currently required for DRL applications. The integration of hierarchical reinforcement learning, where high-level policies set ESG targets and low-level policies execute trades, offers another promising architectural direction.

The integration of diverse data modalities offers significant opportunities for more comprehensive and accurate ESG assessment. Satellite imagery can provide objective measures of environmental impact, from deforestation monitoring to pollution detection, complementing self-reported corporate disclosures. Audio and video analysis of earnings calls and shareholder meetings can detect ESG-related sentiment and commitment that may not appear in transcripts. Social media monitoring across multiple platforms and languages can provide real-time indicators of emerging ESG controversies. Developing fusion architectures that effectively integrate these heterogeneous data sources while handling missing modalities and varying update frequencies represents a substantial technical challenge with high potential impact.

Improving model interpretability is critical for broader institutional adoption of AI in sustainable finance, and this area deserves sustained research attention. ESG-specific explanation methods should highlight which sustainability factors drive predictions and recommendations, enabling analysts to validate AI outputs against their domain expertise. Visualization tools that make AI decisions accessible to non-technical stakeholders, including clients and board members, would facilitate trust and adoption. Moving beyond correlation-based analysis toward causal inference methods that can identify true cause-and-effect relationships between ESG factors and outcomes would represent a major methodological advancement. The development of inherently interpretable models that sacrifice minimal performance compared to black-box alternatives deserves investigation.

Real-time ESG monitoring capabilities would transform sustainable finance from a largely backward-looking to a forward-looking discipline. Streaming data architectures that continuously ingest news, social media, regulatory filings, and alternative data sources could enable immediate detection of ESG-relevant events. Early warning systems for ESG controversies and emerging risks could provide valuable lead time for portfolio adjustments. Adaptive models that can incorporate new ESG-relevant information within hours rather than waiting for quarterly rating updates would better serve investors in fast-moving markets.

Federated learning \citep{mcmahan2017} and privacy-preserving techniques could address significant data sharing barriers in sustainable finance. Many organizations possess valuable proprietary ESG data but cannot share it due to competitive or regulatory concerns. Federated learning frameworks that allow collaborative model training without centralizing sensitive data could unlock this distributed knowledge. Privacy-preserving benchmarking techniques could enable companies to compare their ESG performance against peers without revealing confidential details. Differential privacy approaches could allow ESG data analysis while providing mathematical guarantees against information leakage.

Climate-related financial risk assessment represents a growing priority area where AI can make substantial contributions \citep{battiston2017}. Physical risk models that integrate climate projections with asset-level geographic data could quantify exposure to extreme weather, sea level rise, and other climate impacts. Transition risk models that assess company vulnerability to policy changes, technology shifts, and market preferences in the decarbonization transition require sophisticated scenario analysis capabilities. Long-horizon prediction models that extend beyond typical financial forecasting timeframes are needed to capture climate risks that may materialize over decades. The integration of climate science models with financial AI systems represents a challenging interdisciplinary frontier.

Finally, the environmental footprint of AI itself is emerging as a research priority for sustainable finance. \citet{elbouknify2026} introduce Green AI principles for finance and propose model design choices that reduce energy consumption and computational waste, framing responsible adoption of AI as a sustainability objective in its own right. Future work in this direction should embed Green AI considerations such as energy efficiency, carbon-aware training, and model compression into the evaluation of AI systems for ESG and sustainable finance applications, ensuring that the means by which the field pursues sustainability do not themselves undermine it.

\section{Conclusion}
\label{sec:conclusion}

This review has provided a comprehensive overview of AI applications in sustainable finance, presenting a structured taxonomy and their specific uses in ESG-related investment decision-making and a framework on AI-powered sustainable finance, that systematizes AI applications to bridge ESG data weaknesses.. 

Our review reveals that while significant progress has been made, the field remains in a dynamic developmental stage with substantial room for advancement. Supervised learning approaches, particularly ensemble methods and deep learning, have been successfully applied to ESG score prediction, financial distress forecasting, and controversy detection, demonstrating the value of AI for ESG assessment tasks. Deep reinforcement learning has emerged as a particularly promising approach for ESG-integrated portfolio management, with recent work demonstrating both performance benefits from ESG-aware policies and methods for explaining agent behavior. Natural language processing, especially transformer-based models, has transformed the analysis of sustainability reports, news, and social media, enabling systematic processing of the vast textual information relevant to ESG assessment. Optimization algorithms including Bayesian optimization and genetic algorithms have enabled more sophisticated ESG-integrated portfolio construction that balances multiple objectives.

However, substantial challenges persist that the research community and industry must address. Data quality issues, including missing scores, rating disagreements across providers, and temporal limitations, continue to hinder AI applications and require both methodological solutions and industry standardization efforts. Model interpretability remains a concern, particularly as regulatory requirements for transparency increase and institutional investors demand explainable recommendations. The dynamic nature of greenwashing and the lack of standardized ESG metrics further complicate the development of robust AI systems that can distinguish genuine sustainability from marketing rhetoric.

The first contribution of this paper lies in the development of a comprehensive taxonomy of AI approaches applicable to sustainable finance. By organizing the literature around three main AI paradigms—machine learning, natural language processing, and optimization—this taxonomy provides a coherent structure that clarifies methodological differences and complementarities across studies. In contrast to prior reviews that focus on specific techniques or narrow domains, the proposed taxonomy offers an integrated view of how AI models are used to support ESG assessment, risk prediction, portfolio construction, and monitoring tasks. This structured classification facilitates comparison across studies and helps identify underexplored combinations of AI techniques and ESG applications.

The second contribution is the proposed framework linking AI technologies to core ESG data limitations that hinder sustainable investment decisions. Building on documented challenges in ESG data—such as limited coverage, credibility concerns, uncertainty, unstructured information, lack of parsimony, and inconsistency across providers \citep{chatterji2016, berg2022aggregate, lyon2015}—the framework makes explicit how different AI approaches can mitigate specific data-related barriers. Rather than treating AI as a generic solution, the framework highlights the conditions under which particular AI applications are most effective, thereby improving the alignment between technological choices and sustainability-related decision needs.

A further contribution of this review is to shift the focus from individual AI applications to the systemic role of AI in sustainable finance, potentially strengthening ESG disclosure \citep{asif2023}. The analysis shows that AI’s primary value lies in enhancing the usability and decision relevance of imperfect ESG data, rather than in generating entirely new sustainability metrics. By integrating heterogeneous data sources, extracting signals from unstructured disclosures, and supporting multi-objective decision-making, AI technologies act as enabling mechanisms that complement, rather than replace, regulatory frameworks and ESG standards \citep{oecd2020, bis2023}. This perspective helps reconcile the growing interest in AI-driven solutions with persistent concerns about ESG data quality and standardization.

This study has limitations typical of integrative literature reviews. The scope and final sample are necessarily selective rather than exhaustive, which may limit the generalizability of the findings across ESG domains and AI applications. In addition, the rapid evolution of AI technologies means that some approaches discussed may quickly become outdated, highlighting the need for ongoing updates to the proposed taxonomy and framework. Nonetheless, the findings of this study are highly relevant to researchers, practitioners, and policymakers seeking to understand how different AI technologies can be systematically aligned with ESG data challenges to support more effective sustainable investment decision-making. 

Looking forward, we have identified several promising research directions including advanced reinforcement learning architectures for portfolio management, multimodal AI systems for comprehensive ESG assessment, explainable AI methods tailored for sustainable finance, real-time ESG monitoring capabilities, federated learning for privacy-preserving collaboration, and AI integration with climate science for risk assessment. As sustainable finance continues to grow in importance with increasing regulatory mandates and investor demand, and as AI technologies continue to advance, the intersection of these fields offers tremendous potential for positive societal impact. 
 
\section*{Acknowledgments}

This work was partially supported by the Madrid Regional Government (Grant PHS-2024/PH-HUM-294), Spain) and CaixaBank AM.

\bibliographystyle{apalike}

\begin{thebibliography}{}

\bibitem[Agosto et~al., 2025]{agosto2025}
Agosto, A., Cerchiello, P., and Giudici, P. (2025).
\newblock {Bayesian} learning models to measure the relative impact of {ESG}
  factors on credit ratings.
\newblock {\em International Journal of Data Science and Analytics},
  20(2):357--368.

\bibitem[Agosto et~al., 2023]{agosto2023}
Agosto, A., Giudici, P., and Tanda, A. (2023).
\newblock How to combine {ESG} scores? a proposal based on credit rating
  prediction.
\newblock {\em Corporate Social Responsibility and Environmental Management},
  30(6):3222--3230.

\bibitem[Al-Sartawi et~al., 2022]{alsartawi2022}
Al-Sartawi, A. M.~A., Hussainey, K., and Razzaque, A. (2022).
\newblock The role of artificial intelligence in sustainable finance.
\newblock {\em Journal of Sustainable Finance \& Investment}, pages 1--6.

\bibitem[Al-Sartawi et~al., 2024]{alsartawi2023}
Al-Sartawi, A. M. A.~M., Abd~Wahab, M.~H., and Hussainey, K., editors (2024).
\newblock {\em Global Economic Revolutions: Big Data Governance and Business
  Analytics for Sustainability}, Communications in Computer and Information
  Science, Cham. Springer Nature Switzerland.

\bibitem[Asif et~al., 2023]{asif2023}
Asif, M., Searcy, C., and Castka, P. (2023).
\newblock {ESG} and {Industry 5.0}: The role of technologies in enhancing {ESG}
  disclosure.
\newblock {\em Technological Forecasting and Social Change}, 195:122806.

\bibitem[Bang et~al., 2023]{bang2023}
Bang, J., Ryu, D., and Yu, J. (2023).
\newblock {ESG} controversies and investor trading behavior in the {Korean}
  market.
\newblock {\em Finance Research Letters}, 54:103750.

\bibitem[{Bank for International Settlements}, 2023]{bis2023}
{Bank for International Settlements} (2023).
\newblock Annual economic report 2023.
\newblock Technical report, Bank for International Settlements, Basel.

\bibitem[Battiston et~al., 2017]{battiston2017}
Battiston, S., Mandel, A., Monasterolo, I., Sch{\"u}tze, F., and Visentin, G.
  (2017).
\newblock A climate stress-test of the financial system.
\newblock {\em Nature Climate Change}, 7(4):283--288.

\bibitem[Ben~Jabeur et~al., 2021]{jabeur2021}
Ben~Jabeur, S., Khalfaoui, R., and Ben~Arfi, W. (2021).
\newblock The effect of green energy, global environmental indexes, and stock
  markets in predicting oil price crashes: Evidence from explainable machine
  learning.
\newblock {\em Journal of Environmental Management}, 298:113511.

\bibitem[Berg et~al., 2022]{berg2022aggregate}
Berg, F., K{\"o}lbel, J.~F., and Rigobon, R. (2022).
\newblock Aggregate confusion: The divergence of {ESG} ratings.
\newblock {\em Review of Finance}, 26(6):1315--1344.

\bibitem[Bingler et~al., 2022]{bingler2022}
Bingler, J.~A., Kraus, M., Leippold, M., and Webersinke, N. (2022).
\newblock Cheap talk and cherry-picking: What {ClimateBert} has to say on
  corporate climate risk disclosures.
\newblock {\em Finance Research Letters}, 47:102776.

\bibitem[Bishop, 2006]{bishop2006}
Bishop, C.~M. (2006).
\newblock {\em Pattern Recognition and Machine Learning}.
\newblock Springer.

\bibitem[Boiral and Henri, 2017]{boiral2017}
Boiral, O. and Henri, J.-F. (2017).
\newblock Is sustainability performance comparable? {A} study of {GRI} reports
  of mining organizations.
\newblock {\em Business \& Society}, 56(2):283--317.

\bibitem[Bolton and Kacperczyk, 2021]{bolton2021}
Bolton, P. and Kacperczyk, M. (2021).
\newblock Do investors care about carbon risk?
\newblock {\em Journal of Financial Economics}, 142(2):517--549.

\bibitem[Burnaev et~al., 2023]{burnaev2023}
Burnaev, E., Mironov, E., Shpilman, A., Mironenko, M., and Katalevsky, D.
  (2023).
\newblock Practical {AI} cases for solving {ESG} challenges.
\newblock {\em Sustainability}, 15(17):12731.

\bibitem[Camilleri, 2015]{camilleri2015}
Camilleri, M.~A. (2015).
\newblock Environmental, social and governance disclosures in {Europe}.
\newblock {\em Sustainability Accounting, Management and Policy Journal},
  6(2):224--242.

\bibitem[Campiglio et~al., 2025]{campiglio2025}
Campiglio, E., Deyris, J., Romelli, D., and Scalisi, G. (2025).
\newblock Warning words in a warming world: Central bank communication and
  climate change.
\newblock {\em European Economic Review}, 178:105101.

\bibitem[Chatterji et~al., 2016]{chatterji2016}
Chatterji, A.~K., Durand, R., Levine, D.~I., and Touboul, S. (2016).
\newblock Do ratings of firms converge? {Implications} for managers, investors
  and strategy researchers.
\newblock {\em Strategic Management Journal}, 37(8):1597--1614.

\bibitem[Chatterji et~al., 2009]{chatterji2009}
Chatterji, A.~K., Levine, D.~I., and Toffel, M.~W. (2009).
\newblock How well do social ratings actually measure corporate social
  responsibility?
\newblock {\em Journal of Economics \& Management Strategy}, 18(1):125--169.

\bibitem[Citterio and King, 2023]{citterio2023}
Citterio, A. and King, T. (2023).
\newblock The role of environmental, social, and governance ({ESG}) in
  predicting bank financial distress.
\newblock {\em Finance Research Letters}, 51:103411.

\bibitem[Cooper, 1988]{cooper1988}
Cooper, H.~M. (1988).
\newblock Organizing knowledge syntheses: A taxonomy of literature reviews.
\newblock {\em Knowledge in Society}, 1(1):104--126.

\bibitem[D'Amato et~al., 2021]{damato2021}
D'Amato, V., D'Ecclesia, R., and Levantesi, S. (2021).
\newblock Fundamental ratios as predictors of {ESG} scores: A machine learning
  approach.
\newblock {\em Decisions in Economics and Finance}, 44(2):1087--1110.

\bibitem[Day et~al., 2023]{day2023}
Day, M.-Y., Yang, C.-Y., and Ni, Y. (2023).
\newblock Portfolio dynamic trading strategies using deep reinforcement
  learning.
\newblock {\em Soft Computing}, 27(15-16):8715--8730.

\bibitem[de-la Rica-Escudero et~al., 2025]{delarica2025}
de-la Rica-Escudero, A., Garrido-Merch{\'a}n, E.~C., and Coronado-Vaca, M.
  (2025).
\newblock Explainable post hoc portfolio management financial policy of a deep
  reinforcement learning agent.
\newblock {\em PLOS ONE}, 20(1):e0315528.

\bibitem[De~Lucia et~al., 2020]{delucia2020}
De~Lucia, C., Pazienza, P., and Bartlett, M. (2020).
\newblock Does good {ESG} lead to better financial performances by firms?
  machine learning and logistic regression models of public enterprises in
  {Europe}.
\newblock {\em Sustainability}, 12(13):5317.

\bibitem[Devlin et~al., 2019]{devlin2019}
Devlin, J., Chang, M.-W., Lee, K., and Toutanova, K. (2019).
\newblock {BERT}: Pre-training of deep bidirectional transformers for language
  understanding.
\newblock In {\em Proceedings of the 2019 Conference of the North American
  Chapter of the Association for Computational Linguistics: Human Language
  Technologies}, pages 4171--4186.

\bibitem[Eccles et~al., 2014]{eccles2014}
Eccles, R.~G., Ioannou, I., and Serafeim, G. (2014).
\newblock The impact of corporate sustainability on organizational processes
  and performance.
\newblock {\em Management Science}, 60(11):2835--2857.

\bibitem[Elbouknify et~al., 2026]{elbouknify2026}
Elbouknify, I., Machado, M.~R., and Iannario, M. (2026).
\newblock Designing green artificial intelligence ({Green AI}) models for
  finance: A novel approach for sustainable and responsible adoption.
\newblock {\em Financial Innovation}, 12(1):96.

\bibitem[Eti et~al., 2026]{eti2026}
Eti, S., Y{\"u}ksel, S., Din{\c{c}}er, H., G{\"o}kalp, Y., Y{\i}ld{\i}z, H.,
  Erg{\"u}n, E., and Acar, M. (2026).
\newblock Machine learning-enhanced fuzzy framework for prioritizing financial
  risk management techniques in renewable energy projects.
\newblock {\em Financial Innovation}, 12(1):85.

\bibitem[Friede et~al., 2015]{friede2015esg}
Friede, G., Busch, T., and Bassen, A. (2015).
\newblock {ESG} and financial performance: Aggregated evidence from more than
  2000 empirical studies.
\newblock {\em Journal of Sustainable Finance \& Investment}, 5(4):210--233.

\bibitem[Garc{\'\i}a~Garc{\'\i}a et~al., 2022]{garcia2022}
Garc{\'\i}a~Garc{\'\i}a, F., Gankova-Ivanova, T., Gonz{\'a}lez-Bueno, J.,
  Oliver-Muncharaz, J., and Tamosiuniene, R. (2022).
\newblock What is the cost of maximizing {ESG} performance in the portfolio
  selection strategy? the case of the dow jones index average stocks.
\newblock {\em Entrepreneurship and Sustainability Issues}, 9(4):178--192.

\bibitem[Garrido-Merch{\'a}n et~al., 2023a]{garrido2023climatebert}
Garrido-Merch{\'a}n, E.~C., Gonz{\'a}lez-Barthe, C., and Coronado~Vaca, M.
  (2023a).
\newblock Fine-tuning {ClimateBert} transformer with {ClimaText} for the
  disclosure analysis of climate-related financial risks.
\newblock {\em arXiv preprint arXiv:2303.13373}.

\bibitem[Garrido-Merch{\'a}n et~al., 2023b]{garrido2023bo}
Garrido-Merch{\'a}n, E.~C., Gonz{\'a}lez~Piris, G., and Coronado~Vaca, M.
  (2023b).
\newblock Bayesian optimization of {ESG} (environmental social governance)
  financial investments.
\newblock {\em Environmental Research Communications}, 5(5):055003.

\bibitem[Garrido-Merch{\'a}n et~al., 2024]{garrido2024mobo}
Garrido-Merch{\'a}n, E.~C., Mora-Figueroa, S., and Coronado-Vaca, M. (2024).
\newblock Multi-objective {Bayesian} optimization of deep reinforcement
  learning for environmental, social, and governance ({ESG}) financial
  portfolio management.
\newblock {\em Intelligent Systems in Accounting, Finance and Management},
  31(4):e1576.

\bibitem[Garrido-Merch{\'a}n et~al., 2023c]{garrido2023drl}
Garrido-Merch{\'a}n, E.~C., Mora-Figueroa-Cruz-Guzm{\'a}n, S., and
  Coronado-Vaca, M. (2023c).
\newblock Deep reinforcement learning for {ESG} financial portfolio management.
\newblock {\em arXiv preprint arXiv:2307.09631}.

\bibitem[Gillan et~al., 2021]{gillan2021}
Gillan, S.~L., Koch, A., and Starks, L.~T. (2021).
\newblock Firms and social responsibility: A review of {ESG} and {CSR} research
  in corporate finance.
\newblock {\em Journal of Corporate Finance}, 66:101889.

\bibitem[{Global Sustainable Investment Alliance}, 2021]{gsia2021}
{Global Sustainable Investment Alliance} (2021).
\newblock Global sustainable investment review 2020.
\newblock Technical report, Global Sustainable Investment Alliance.

\bibitem[{Global Sustainable Investment Alliance}, 2022]{gsia2022}
{Global Sustainable Investment Alliance} (2022).
\newblock Global sustainable investment review 2022.
\newblock Technical report, GSIA.

\bibitem[Goodell et~al., 2021]{goodell2021}
Goodell, J.~W., Kumar, S., Lim, W.~M., and Pattnaik, D. (2021).
\newblock Artificial intelligence and machine learning in finance: Identifying
  foundations, themes, and research clusters from bibliometric analysis.
\newblock {\em Journal of Behavioral and Experimental Finance}, 32:100577.

\bibitem[Goodfellow et~al., 2016]{goodfellow2016}
Goodfellow, I., Bengio, Y., and Courville, A. (2016).
\newblock {\em Deep Learning}.
\newblock MIT Press.

\bibitem[Gozalo-Brizuela and Garrido-Merch{\'a}n, 2023]{gozalo2023}
Gozalo-Brizuela, R. and Garrido-Merch{\'a}n, E.~C. (2023).
\newblock {ChatGPT} is not all you need. a state of the art review of large
  generative {AI} models.
\newblock {\em arXiv preprint arXiv:2301.04655}.

\bibitem[Gupta et~al., 2024]{gupta2024}
Gupta, A., Chadha, A., and Tewari, V. (2024).
\newblock A natural language processing model on {BERT} and {YAKE} technique
  for keyword extraction on sustainability reports.
\newblock {\em IEEE Access}.

\bibitem[Gupta and Yan, 2025]{guptayan2025}
Gupta, S. and Yan, H. (2025).
\newblock Using large language models to estimate novel risk: Impact on
  volatility.
\newblock {\em Journal of Portfolio Management}, 51(7).

\bibitem[Hastie et~al., 2009]{hastie2009}
Hastie, T., Tibshirani, R., and Friedman, J. (2009).
\newblock {\em The Elements of Statistical Learning}.
\newblock Springer.

\bibitem[Jaiswal et~al., 2024]{jaiswal2024}
Jaiswal, R., Gupta, S., and Tiwari, A.~K. (2024).
\newblock Decoding mood of the twitterverse on {ESG} investing: Opinion mining
  and key themes using machine learning.
\newblock {\em Management Research Review}, 47(8):1221--1252.

\bibitem[James et~al., 2023]{james2023}
James, G., Witten, D., Hastie, T., Tibshirani, R., and Taylor, J. (2023).
\newblock {\em An Introduction to Statistical Learning}.
\newblock Springer, 2nd edition.

\bibitem[Kazakov et~al., 2023]{kazakov2023}
Kazakov, A., Denisova, S., Barsola, I., Kalugina, E., Molchanova, I., Egorov,
  I., Kosterina, A., Tereshchenko, E., Shutikhina, L., Doroshchenko, I.,
  Sotiriadi, N., and Budennyy, S. (2023).
\newblock {ESGify}: Automated classification of environmental, social and
  corporate governance risks.
\newblock {\em Doklady Mathematics}, 108(Suppl 2):S529--S540.

\bibitem[K{\"o}lbel et~al., 2020]{kolbel2020}
K{\"o}lbel, J.~F., Heeb, F., Paetzold, F., and Busch, T. (2020).
\newblock Can sustainable investing save the world? {R}eviewing the mechanisms
  of investor impact.
\newblock {\em Organization \& Environment}, 33(4):554--574.

\bibitem[LeCun et~al., 2015]{lecun2015}
LeCun, Y., Bengio, Y., and Hinton, G. (2015).
\newblock Deep learning.
\newblock {\em Nature}, 521(7553):436--444.

\bibitem[Lee et~al., 2024]{lee2024}
Lee, H., Lee, S.~H., Park, H., Kim, J.~H., and Jung, H.~S. (2024).
\newblock {ESG2PreEM}: Automated {ESG} grade assessment framework using
  pre-trained ensemble models.
\newblock {\em Heliyon}, 10(4):e26404.

\bibitem[Lheureux, 2023]{lheureux2023}
Lheureux, Y. (2023).
\newblock Predictive insights: Leveraging {Twitter} sentiments and machine
  learning for environmental, social and governance controversy prediction.
\newblock {\em Journal of Computational Social Science}, 7:23--44.

\bibitem[Liao, 2025]{liao2025}
Liao, C.-F. (2025).
\newblock {ESG} disclosure frequency and its association with market
  performance: Evidence from {Taiwan}.
\newblock {\em Sustainability}, 17(17):7812.

\bibitem[Lim et~al., 2021]{lim2021}
Lim, W.~M., Yap, S.-F., and Makkar, M. (2021).
\newblock Home sharing in marketing and tourism at a tipping point: What do we
  know, how do we know, and where should we be heading?
\newblock {\em Journal of Business Research}, 122:534--566.

\bibitem[Liu et~al., 2020]{liu2020finrl}
Liu, X.-Y., Yang, H., Chen, Q., Zhang, R., Yang, L., Xiao, B., and Wang, C.~D.
  (2020).
\newblock {FinRL}: A deep reinforcement learning library for automated stock
  trading in quantitative finance.
\newblock {\em arXiv preprint arXiv:2011.09607}.

\bibitem[Loughran and McDonald, 2016]{loughran2016}
Loughran, T. and McDonald, B. (2016).
\newblock Textual analysis in accounting and finance: A survey.
\newblock {\em Journal of Accounting Research}, 54(4):1187--1230.

\bibitem[Lundberg and Lee, 2017]{lundberg2017}
Lundberg, S.~M. and Lee, S.-I. (2017).
\newblock A unified approach to interpreting model predictions.
\newblock In {\em Advances in Neural Information Processing Systems},
  volume~30, pages 4766--4777.

\bibitem[Lyon and Montgomery, 2015]{lyon2015}
Lyon, T.~P. and Montgomery, A.~W. (2015).
\newblock The means and end of greenwash.
\newblock {\em Organization \& Environment}, 28(2):223--249.

\bibitem[Malik et~al., 2025]{malik2025}
Malik, M., Mamun, K., and Osman, S. M.~I. (2025).
\newblock Does corruption control enhance {ESG}-induced firm value? insights
  from machine learning analysis.
\newblock {\em Finance Research Letters}, 72:106572.

\bibitem[Maree and Omlin, 2022]{maree2022}
Maree, C. and Omlin, C.~W. (2022).
\newblock Balancing profit, risk, and sustainability for portfolio management.
\newblock In {\em 2022 IEEE Symposium on Computational Intelligence for
  Financial Engineering and Economics (CIFEr)}, pages 1--8. IEEE.

\bibitem[Markowitz, 1952]{markowitz1952}
Markowitz, H. (1952).
\newblock Portfolio selection.
\newblock {\em The Journal of Finance}, 7(1):77--91.

\bibitem[Martin-Melero et~al., 2025]{martinmelero2025}
Martin-Melero, I., Gomez-Martinez, R., Medrano-Garcia, M.~L., and
  Hernandez-Perlines, F. (2025).
\newblock Comparison of sectorial and financial data for {ESG} scoring of
  mutual funds with machine learning.
\newblock {\em Financial Innovation}, 11(1):84.

\bibitem[McMahan et~al., 2017]{mcmahan2017}
McMahan, B., Moore, E., Ramage, D., Hampson, S., and Aguera~y Arcas, B. (2017).
\newblock Communication-efficient learning of deep networks from decentralized
  data.
\newblock In {\em Proceedings of the 20th International Conference on
  Artificial Intelligence and Statistics}, pages 1273--1282.

\bibitem[Meng et~al., 2022]{meng2022}
Meng, T., Yahya, M.~H., and Chai, J. (2022).
\newblock Deep learning model for stock excess return prediction based on
  nonlinear random matrix and {ESG} factor.
\newblock {\em Mathematical Problems in Engineering}, 2022:5239493.

\bibitem[Mnih et~al., 2015]{mnih2015}
Mnih, V., Kavukcuoglu, K., Silver, D., Rusu, A.~A., Veness, J., Bellemare,
  M.~G., Graves, A., Riedmiller, M., Fidjeland, A.~K., Ostrovski, G., Petersen,
  S., Beattie, C., Sadik, A., Antonoglou, I., King, H., Kumaran, D., Wierstra,
  D., Legg, S., and Hassabis, D. (2015).
\newblock Human-level control through deep reinforcement learning.
\newblock {\em Nature}, 518(7540):529--533.

\bibitem[Molitor et~al., 2023]{raghupathi2023}
Molitor, D., Raghupathi, V., and Saharia, A. (2023).
\newblock Identifying key issues in climate change litigation: A machine
  learning text analytic approach.
\newblock {\em Sustainability}, 15(23):16530.

\bibitem[{OECD}, 2020]{oecd2020}
{OECD} (2020).
\newblock Developing sustainable finance definitions and taxonomies.
\newblock Technical report, OECD Publishing, Paris.

\bibitem[Pankratz et~al., 2023]{pankratz2023}
Pankratz, N., Bauer, R., and Derwall, J. (2023).
\newblock Climate change, firm performance, and investor surprises.
\newblock {\em Management Science}, 69(12):7352--7398.

\bibitem[Paul and Benito, 2018]{paulbenito2018}
Paul, J. and Benito, G. R.~G. (2018).
\newblock A review of research on outward foreign direct investment from
  emerging countries, including {China}: what do we know, how do we know and
  where should we be heading?
\newblock {\em Asia Pacific Business Review}, 24(1):90--115.

\bibitem[Paul and Criado, 2020]{paul2020}
Paul, J. and Criado, A.~R. (2020).
\newblock The art of writing literature review: What do we know and what do we
  need to know?
\newblock {\em International Business Review}, 29(4):101717.

\bibitem[Paul et~al., 2021]{paul2021spar}
Paul, J., Lim, W.~M., O'Cass, A., Hao, A.~W., and Bresciani, S. (2021).
\newblock Scientific procedures and rationales for systematic literature
  reviews ({SPAR-4-SLR}).
\newblock {\em International Journal of Consumer Studies}, 45(4):O1--O16.

\bibitem[Pikatza-Gorrotxategi et~al., 2024]{pikatza2024}
Pikatza-Gorrotxategi, N., Borregan-Alvarado, J., and Alvarez-Meaza, I. (2024).
\newblock News and {ESG} investment criteria: What's behind it?
\newblock {\em Social Network Analysis and Mining}, 14(1):47.

\bibitem[Porter and van~der Linde, 1995]{porter1995}
Porter, M.~E. and van~der Linde, C. (1995).
\newblock Toward a new conception of the environment-competitiveness
  relationship.
\newblock {\em Journal of Economic Perspectives}, 9(4):97--118.

\bibitem[Rajan and Zingales, 2003]{rajan2003}
Rajan, R.~G. and Zingales, L. (2003).
\newblock The great reversals: the politics of financial development in the
  twentieth century.
\newblock {\em Journal of Financial Economics}, 69(1):5--50.

\bibitem[Redondo and Aracil, 2024]{redondo2024}
Redondo, H. and Aracil, E. (2024).
\newblock Climate-related credit risk: Rethinking the credit risk framework.
\newblock {\em Global Policy}, 15(S1):21--33.

\bibitem[Ribeiro et~al., 2016]{ribeiro2016}
Ribeiro, M.~T., Singh, S., and Guestrin, C. (2016).
\newblock ``{Why} should {I} trust you?'': Explaining the predictions of any
  classifier.
\newblock In {\em Proceedings of the 22nd ACM SIGKDD International Conference
  on Knowledge Discovery and Data Mining}, pages 1135--1144.

\bibitem[Russell and Norvig, 2010]{russell2010}
Russell, S. and Norvig, P. (2010).
\newblock {\em Artificial Intelligence: A Modern Approach}.
\newblock Prentice Hall, 3rd edition.

\bibitem[Schoenmaker, 2017]{schoenmaker2017}
Schoenmaker, D. (2017).
\newblock Investing for the common good: A sustainable finance framework.
\newblock Technical report, Bruegel, Brussels.

\bibitem[Schoenmaker and Schramade, 2019]{schoenmaker2019}
Schoenmaker, D. and Schramade, W. (2019).
\newblock Investing for long-term value creation.
\newblock {\em Journal of Sustainable Finance \& Investment}, 9(4):356--377.

\bibitem[Schulman et~al., 2017]{schulman2017}
Schulman, J., Wolski, F., Dhariwal, P., Radford, A., and Klimov, O. (2017).
\newblock Proximal policy optimization algorithms.
\newblock {\em arXiv preprint arXiv:1707.06347}.

\bibitem[Seles et~al., 2018]{seles2018}
Seles, B. M. R.~P., de~Sousa~Jabbour, A. B.~L., Jabbour, C. J.~C., and
  de~Camargo~Fiorini, P. (2018).
\newblock Business opportunities and challenges as the two sides of the climate
  change: Corporate responses and potential implications for big data
  management towards a low carbon society.
\newblock {\em Journal of Cleaner Production}, 189:763--774.

\bibitem[Shahriari et~al., 2016]{shahriari2016}
Shahriari, B., Swersky, K., Wang, Z., Adams, R.~P., and De~Freitas, N. (2016).
\newblock Taking the human out of the loop: A review of {Bayesian}
  optimization.
\newblock {\em Proceedings of the IEEE}, 104(1):148--175.

\bibitem[Singhania et~al., 2025]{singhania2025}
Singhania, M., Chadha, G., and {Anisha} (2025).
\newblock Sustainability accounting research over three decades: A
  scientometric meta-analysis.
\newblock {\em Corporate Social Responsibility and Environmental Management},
  32(2):1698--1734.

\bibitem[Snyder, 2019]{snyder2019}
Snyder, H. (2019).
\newblock Literature review as a research methodology: An overview and
  guidelines.
\newblock {\em Journal of Business Research}, 104:333--339.

\bibitem[Song et~al., 2025]{song2025}
Song, J., Huang, J., Peng, X., Sumran, A., and Wang, Y. (2025).
\newblock Unveiling internal drivers of corporate greenwashing: A machine
  learning approach.
\newblock {\em Journal of Environmental Planning and Management}, pages 1--23.

\bibitem[Souto and Moradi, 2026]{souto2026}
Souto, H.~G. and Moradi, A. (2026).
\newblock Enhancing financial risk management: A novel multivariate neural
  network approach for realized covariance matrix prediction.
\newblock {\em Financial Innovation}, 12(1):25.

\bibitem[Sutton and Barto, 2018]{sutton2018}
Sutton, R.~S. and Barto, A.~G. (2018).
\newblock {\em Reinforcement Learning: An Introduction}.
\newblock MIT Press, 2nd edition.

\bibitem[Svanberg et~al., 2022]{svanberg2022}
Svanberg, J., Ardeshiri, T., Samsten, I., {\"O}hman, P., Neidermeyer, P.~E.,
  Rana, T., and Danielson, M. (2022).
\newblock Corporate governance performance ratings with machine learning.
\newblock {\em Intelligent Systems in Accounting, Finance and Management},
  29(1):50--68.

\bibitem[{Task Force on Climate-related Financial Disclosures}, 2017]{tcfd2017}
{Task Force on Climate-related Financial Disclosures} (2017).
\newblock Recommendations of the {Task Force on Climate-related Financial
  Disclosures}.
\newblock Technical report, Financial Stability Board.

\bibitem[Teoh et~al., 2019]{teoh2019}
Teoh, T.~T., Heng, Q.~K., Chia, J.~J., Shie, J.~M., Liaw, S.~W., Yang, M., and
  Nguwi, Y.~Y. (2019).
\newblock Machine learning-based corporate social responsibility prediction.
\newblock In {\em 2019 IEEE International Conference on Cybernetics and
  Intelligent Systems (CIS) and IEEE Conference on Robotics, Automation and
  Mechatronics (RAM)}, pages 501--505. IEEE.

\bibitem[Torraco, 2016]{torraco2016}
Torraco, R.~J. (2016).
\newblock Writing integrative literature reviews: Using the past and present to
  explore the future.
\newblock {\em Human Resource Development Review}, 15(4):404--428.

\bibitem[{United Nations}, 2012]{un2012}
{United Nations} (2012).
\newblock The future we want: Outcome document of the {United Nations}
  conference on sustainable development.
\newblock Technical report, United Nations, Rio de Janeiro.

\bibitem[{United Nations}, 2015]{un2015}
{United Nations} (2015).
\newblock Transforming our world: The 2030 agenda for sustainable development.
\newblock Technical Report A/RES/70/1, United Nations General Assembly.

\bibitem[{United Nations Conference on Trade and Development},
  2023]{unctad2023}
{United Nations Conference on Trade and Development} (2023).
\newblock World investment report 2023: Investing in sustainable energy for
  all.
\newblock Technical report, UNCTAD, Geneva.

\bibitem[{United Nations Environment Programme}, 2016]{unep2016}
{United Nations Environment Programme} (2016).
\newblock The financial system we need: From momentum to transformation.
\newblock Technical report, UNEP Inquiry into the Design of a Sustainable
  Financial System, Geneva.

\bibitem[{United Nations Framework Convention on Climate Change},
  2015]{unfccc2015}
{United Nations Framework Convention on Climate Change} (2015).
\newblock Adoption of the {Paris Agreement}.
\newblock Technical Report FCCC/CP/2015/L.9/Rev.1, United Nations.

\bibitem[Vaswani et~al., 2017]{vaswani2017}
Vaswani, A., Shazeer, N., Parmar, N., Uszkoreit, J., Jones, L., Gomez, A.~N.,
  Kaiser, {\L}., and Polosukhin, I. (2017).
\newblock Attention is all you need.
\newblock In {\em Advances in Neural Information Processing Systems},
  volume~30, pages 5998--6008.

\bibitem[Webersinke et~al., 2022]{webersinke2022}
Webersinke, N., Kraus, M., Bingler, J.~A., and Leippold, M. (2022).
\newblock {ClimateBERT}: A pretrained language model for climate-related text.
\newblock In {\em Proceedings of the AAAI 2022 Fall Symposium: The Role of AI
  in Responding to Climate Challenges}.

\bibitem[Webster and Watson, 2002]{webster2002}
Webster, J. and Watson, R.~T. (2002).
\newblock Analyzing the past to prepare for the future: Writing a literature
  review.
\newblock {\em MIS Quarterly}, 26(2):xiii--xxiii.

\bibitem[{World Bank}, 2022]{worldbank2022}
{World Bank} (2022).
\newblock Sovereign climate and nature reporting: Proposal for a risks and
  opportunities disclosure framework.
\newblock Technical report, World Bank, Washington, DC.

\bibitem[Xu, 2020]{xu2020}
Xu, X. (2020).
\newblock Trust and financial inclusion: A cross-country study.
\newblock {\em Finance Research Letters}, 35:101310.

\bibitem[Zupic and {\v{C}}ater, 2015]{zupic2015}
Zupic, I. and {\v{C}}ater, T. (2015).
\newblock Bibliometric methods in management and organization.
\newblock {\em Organizational Research Methods}, 18(3):429--472.

\end{thebibliography}

\end{document}